\documentclass{interact}

\newcommand{\be}{\begin{eqnarray}}
\newcommand{\ee}{\end{eqnarray}}

\newcommand{\Caf}{$^{40}{\rm Ca}^+$}
\newcommand{\Caft}{$^{43}{\rm Ca}^+$}

\newcommand{\Mrads}{\times 10^6\,{\rm rad\,s}^{-1}}
\newcommand{\Shalf}{$^2{\rm S}_{1/2}$}
\newcommand{\Phalf}{$^2{\rm P}_{1/2}$}
\newcommand{\Dthalf}{$^2{\rm D}_{3/2}$}
\newcommand{\ket}[1]{|#1\rangle}
\newcommand{\wtrap}{\omega_s}

\newcommand{\pd}[2]
{ \left( \frac{ \partial {#1}}{\partial {#2}} \right)}

\usepackage{graphicx}
\usepackage{amssymb}

\usepackage{pgfplots}
\usetikzlibrary{pgfplots.groupplots}
\begin{document}
\tikzset{dashdot/.style={dash pattern=on .4pt off 3pt on 4pt off 3pt}}
\tikzset{shortdash/.style={dash pattern=on 3pt off 2pt}}

\title{The effect of atomic response time in the theory of Doppler cooling of trapped ions}

\author{
\name{H. Janacek, A. M. Steane, D. M. Lucas, D. N. Stacey}
\thanks{h.a.janacek@gmail.com}
\affil{Department of Atomic and Laser Physics, Clarendon Laboratory, Parks Road, Oxford OX1 3PU, England.}
}

\maketitle

\begin{abstract}
We describe a simple approach to the problem of incorporating the response time of an atom or ion being Doppler-cooled into the theory of the cooling process.  The system being cooled does not in general respond instantly to the changing laser frequencies it experiences in its rest frame, and this ``dynamic effect" can affect significantly the temperatures attainable.  It is particularly important for trapped ions when there is a slow decay out of the cooling cycle requiring the use of a repumping beam.  We treat the cases of trapped ions with two and three internal states, then apply the theory to $^{40}{\rm Ca}^+$.  For this ion experimental data exist showing the ion to be cold under conditions for which heating is predicted if the dynamic effect is neglected.  The present theory accounts for the observed behaviour. 
\end{abstract}

%\pacs{}


%\begin{pacscode}
%\end{pacscode}

Doppler cooling exploits the fact that an atom or ion counter-propagating with a laser beam and absorbing photons from it is subject to a retarding force \cite{PhysRevLett.24.156,75Hansch,75Wineland,PhysRevA.20.1521,PhysRevA.25.35,PhysRevLett.48.596,PhysRevLett.55.48,PhysRevLett.89.093003,RevModPhys.70.685,RevModPhys.75.281,BkMetcalf}.   The simplest practical system is that of an ion, trapped in a harmonic potential, illuminated by a single laser beam detuned to the red from a resonance 
transition \cite{PhysRevLett.40.1639,PhysRevLett.41.233,0022-3700-17-16-019,PhysRevA.46.2668}.  The ion is more likely to absorb
photons when travelling towards the source than when moving in the opposite sense, so that averaged over a cycle of the motion the momentum of the ion is reduced.   An equilibrium temperature is reached when the cooling is balanced by the heating effects associated with spontaneous emission and the stochastic nature of the absorption process. 

The simplest assumption is that the internal states of the system being cooled remain in equilibrium with the radiation field, i.e., that the system can respond instantaneously to the changing Doppler shifts of the laser beams.  We shall refer to this as the quasi-static approximation.  The spontaneous transition rate on the resonance transition is often fast enough for this approximation to be justified.  However, in some ions there is a significant time-lag in the atomic response which can significantly influence the outcome of the cooling process.  We refer to this phenomenon as the ``dynamic effect".  It can occur in any system, but particularly when there is a weak decay from the resonance level to a third level, so that a repumping laser is required to return the ion to the cooling cycle.  While it is then generally a good approximation to neglect the momentum exchange with the repumping beam---there are many cooling cycles in between the unwanted decays---the longer response time associated with the weaker decay makes it necessary to take the dynamic effect into account.
  
The effect has been demonstrated in experiments on the calcium ions \Caf\ and \Caft\ in a Paul trap \cite{1367-2630-18-2-023043}.  The cooling transition,
4s\Shalf\ -- 4p\Phalf\ at 397nm, has an Einstein $A$-coefficient $A_{397}$ of $1.32\times 10^8\,{\rm s}^{-1}$ 
so that in a trap of frequency of order a few MHz one would not expect serious time-lag effects.  
However, the decay 4p\Phalf\ -- 3d\Dthalf\ at $866\,$nm is much weaker, with $A_{866} = 8.4\times 10^6\,{\rm s}^{-1}$.  Over a large frequency range of the repumping laser, in which the quasi-static theory suggests that the ion will heat indefinitely, we
have observed experimentally that a \Caf\ ion remains cold.  A simple theoretical treatment including the dynamic effect accounts for the observations.  In section \ref{s.th} below we outline such a treatment based on the optical Bloch equations, applying it first to two- and three-state systems (sections \ref{s.two} and \ref{s.three}), then to the particular case of \Caf\ (section \ref{s.ion}).  
In each case we give examples where a configuration of lasers which would cause heating of the motion
according to quasi-static theory in fact produces cooling, or vice versa, and we explain why.

Our discussion adopts a semi-classical approach in which the ion's motional degree of freedom is treated classically.
Consequently the model does not account for details of the quantised motion which are typically important near
the ground state of motion or in the Lamb-Dicke limit. The phenomena
we shall discuss are well handled to first approximation by our approach.
A more fully quantised theory of laser cooling of a trapped ion has been studied 
extensively in the Lamb-Dicke limit, see for example
\cite{0022-3700-17-16-019,Wineland:92,PhysRevA.46.2668,PhysRevA.49.421}. These treatments
have been useful for studying the approach to steady state at the lowest temperatures, but
have not remarked the behaviour discussed here. 

In sections \ref{s.th}, \ref{s.two} and \ref{s.three} which are entirely theoretical we work in angular frequency units, but for the comparison with experiment in section \ref{s.ion} it is more convenient to use MHz.

\section{Theory} \label{s.th}

To illustrate the principles involved we consider the simplest possible case, an ion which undergoes simple harmonic motion
with secular frequency $\wtrap$
in an isotropic trap, subject to a laser beam red-detuned from a resonance transition of wavelength $\lambda$.  Other repumping lasers may also be present.  Let the cooling beam be propagating in the positive $x$-direction.  We take the velocity component $V_x$ of the ion to be positive in this direction also.  
We define the detuning $\Delta_c$ of the cooling laser from resonance to be positive if it is towards higher frequency.  At first we neglect the dynamic effect.   Let $R(V_x)$ be the rate at which resonance photons are absorbed.  Then to a sufficient approximation for an ion at mK temperatures being cooled on a resonance transition we can write
\be
 R\left(V_x\right) = R_0+R' V_x              	          		\label{R}
\ee
where $R'$ is the derivative of $R$ with respect to $V_x$ at $V_x  = 0$.  Then it can be shown that 
the equilibrium temperature $T$ is given by 
%\cite{PhysRevA.20.1521,86Stenholm}
\be
\frac{k_{\rm B}T}{h} =  - \frac{\alpha}{\lambda} \left[ \frac{R_0}{R'} \right]
\label{T}
\ee
where $\alpha$ is a factor of order unity which depends on the particular system; for our purposes we shall take $\alpha = 1$.  With a two-state system, the response curve is a simple Lorentzian, and (\ref{T}) allows us to find $T$ for any $\Delta_c$.  The minimum, $T_{\rm min}$, occurs when the transition is unsaturated, and occurs at $\Delta_c = -(A/2)$.  If we now consider a more complex system, for example when there is a repumping laser, $R'$ can be found by solving the optical Bloch equations (OBE) representing the ion in two equilibrium situations, one with laser frequencies corresponding to $V_x = 0$, the other with $V_x$  slightly displaced from zero.  The OBE (see, for example, \cite{BkLoudon,Lindberg:86,BkMetcalf,PhysRevA.31.3704}) govern the time-dependence of the density matrix describing the ion under specified conditions.  The elements can be expressed in terms of real quantities $x_{ij},\; y_{ij}$ such that the population of a state 
$\ket{i}$ is $x_{ii}$ while the coherence between $\ket{i}$ and $\ket{j}$ is  $x_{ij} + {\rm i} y_{ij}$.  The equations are then of the form
\be
\dot{z}_m = \sum_n c_{mn} z_n    \label{zdot}
\ee
where $z_n$ represents any of the  $x_{ij}$ or $y_{ij}$ and the coefficients $c_{mn}$ are all real.  They contain the information specific to the particular problem, including the laser frequencies, linewidths, polarizations and intensities, the matrix elements for stimulated and spontaneous transitions, and the external magnetic field.  For a system in equilibrium they therefore reduce to a set of simultaneous linear equations which can be solved relatively easily even when there are many states.	     

However, the validity of (\ref{R}) depends on the neglect of the dynamic effect.  In reality, the absorption rate at a given velocity depends on the history of the ion up to that point; for example, at $V_x  = 0$ it will depend on whether the velocity is changing from positive to negative or vice-versa.  $R'$ as defined above ceases to be a useful parameter.  

We therefore adopt a different approach.  Let
\be
			V_x = V_0 \cos \wtrap t.
\ee
Then to a good approximation the value of $R$ will also oscillate, according to
 \be
R = R_0 + V_0 \left( P \cos \wtrap t+ Q \sin \wtrap t \right) 
\label{RPQ}
\ee
Cooling is then due to the in-phase component $P$.  Equation (\ref{T}) is replaced by
\be
\frac{k_{\rm B}T}{h} =  - \frac{\alpha}{\lambda} \left[ \frac{R_0}{P} \right]
\label{TP}
\ee
Thus, as $\wtrap \rightarrow 0$, $Q \rightarrow 0$ also, and $P \rightarrow R'$, i.e., as calculated quasi-statically.  When the dynamic effect is significant, we calculate $P$ and $Q$ using the time-dependent OBE.  However, the assumption of the steady state solution (\ref{RPQ}) enables us to avoid explicitly integrating over time.

We illustrate the method by considering an ion illuminated by two lasers, one on the cooling transition (subscript $c$), the other a repumper (subscript $r$).  Otherwise the treatment is quite general; the levels involved may comprise any number of states, and the ion may be in a magnetic field.  The extension to more (or less) complex systems is straightforward.  The time-dependence all arises from the fact that there are oscillating Doppler shifts 
$\delta\omega$ superposed on the detunings, given by 
\be
\delta \omega_c &=& V_0 k_c \cos \wtrap t   \\
\delta \omega_r &=& \pm V_0 k_r \cos \wtrap t  \label{dwr}
\ee
where $k_c = 2\pi/\lambda_c,\; k_r =2\pi/\lambda_r$.  We take the positive sign in (\ref{dwr}), corresponding to co-propagating lasers, since the experimental data presented in section \ref{s.ion} 
were obtained with this configuration.  Thus, the $c_{mn}$ in (\ref{zdot}) are replaced by
$C_{mn} + a_{mn} \cos \wtrap t$, where the $C_{mn}$ are time-independent and the $a_{mn}$ are zero except for the terms representing detunings, in which case they are either $V_0 k_c$ or $V_0 k_r$.  

We first solve the time-independent equations with the $C_{mn}$, i.e., we use the mean values of the detunings.  Let the solutions for this case be $Z_n$.  Then provided the amplitudes of the variations in the matrix elements are very small compared with unity, we can write
\be
z_n = Z_n + V_0 \left( u_n \cos \wtrap t + v_n \sin \wtrap t \right)
\ee
Thus, we assume that the harmonically oscillating detunings produce similarly oscillating density matrix elements, but with amplitudes and phases to be determined.  Substituting (7), (8) and (9) into (\ref{zdot}) and neglecting second order terms we obtain
\be
u_m + \sum_n D_{mn} v_n &=& 0          \label{ueq} \\
v_m - \sum_n \left[ D_{mn} u_n + b_{mn} Z_n \right] &=& 0 \label{nueq} 
\ee
where
\be
D_{mn} &=& C_{mn} / \wtrap \\
b_{mn} &=& a_{mn} / \wtrap .
\ee
We have in (\ref{ueq}, \ref{nueq}) a set of linear equations with constant coefficients, independent of $V_0$, which can be solved in the usual way to find the $u_n$ and $v_n$, and hence $P$ and $Q$.  In the limit $\wtrap \rightarrow 0$, where dynamic effects are negligible, all the $v_n \rightarrow 0$ and the $u_n$ are the same as those given by the quasi-static theory. 

We note that both theories rely on the velocity of the ion remaining low enough for the basic equations, (\ref{R}) for the quasi-static theory and (\ref{RPQ}) for the dynamic theory, to be good approximations.  While this condition is fulfilled for the temperatures of interest here, refinements to the theory are needed to describe the behaviour of systems at elevated temperatures \cite{HugoThesis,1367-2630-18-2-023043}.

\section{The two-state system}  \label{s.two}

We first apply the analysis to the two-state system: lower state $\ket{1}$, upper state $\ket{2}$, transition wavelength $\lambda_{12}$ (corresponding to a frequency $\omega_{12}$), cooled by a single red-detuned laser beam of frequency $\omega_L$.  Although normally one cools on transitions in which the decay rate is fast enough to render the dynamic effect unimportant, in a tight trap this is not necessarily the case.  We specify the system by the following parameters: Rabi frequency on resonance $\Omega_{12}$, Einstein $A$-coefficient for the transition $A_{21}$, and detuning $\Delta_{12} = \omega_L - \omega_{12}$.  For simplicity we set the laser linewidth to zero since it plays no fundamental role in the process.  Then the equilibrium temperature depends on $u_2$, which characterises the in-phase amplitude of the upper state population.  We take $P = A_{21} u_2$.

\begin{figure} [t]
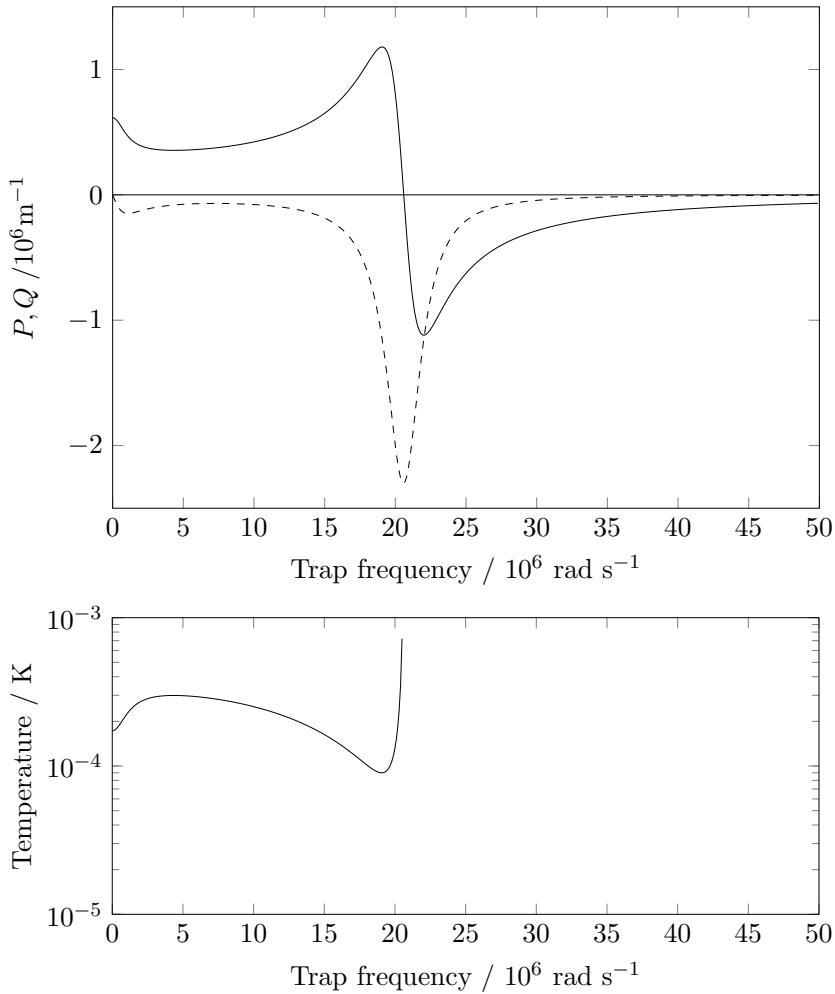

\begin{center}
\hspace{1cm}\begin{tikzpicture}[scale=0.95] \input{fig1} 
%\myBoundingBox
\end{tikzpicture}\\
\rule{1cm}{0pt}\begin{tikzpicture}[scale=0.95] \input{fig1b} 
%\myBoundingBox
\end{tikzpicture}
\end{center}
\caption{Results of the approximate method for solving the Bloch equations for a two-state system.  In the upper plot, the full (dashed) line is the quantity $P\, (Q)$, both defined in equation (\ref{RPQ}).  $P$ and $Q$ give respectively the in-phase and out-of-phase amplitudes of the photon emission rate for an oscillating ion with velocity amplitude $1\,$ms$^{-1}$.  The lower plot shows the equilibrium temperature $T$ from equation (\ref{TP}); $T \sim 1/P$, so the peak value of $P$ corresponds to the lowest temperature attainable.
For $\wtrap > \Omega_{12}'$, where $P$ is negative, the ion will heat.  See text for discussion and the values of the parameters.}
\end{figure}

Figure 1 shows the results for $P$ and $Q$ as functions of $\wtrap$ for $\lambda_{12} = 400\,$nm, $\Omega_{12} = 20\times 10^6\,$rad s$^{-1}$, $A_{21} = 2\times 10^6\,{\rm s}^{-1}$ and 
$\Delta_{12} = -5\times 10^6\,$rad s$^{-1}$.  The red detuning ensures cooling according to the quasi-static theory.  The equilibrium temperature found using (\ref{TP}) is also shown. 

%{\em Note that I am writing these frequencies in a way that I hope my co-authors will find unambiguous, but I don’t care how they appear in the paper.  
%I am putting them all $\times 10^6$ even when logically one would write a different exponent because to have them all the same makes comparison of 
%magnitudes so much easier.  The reason for choosing these high frequencies even when we are discussing the simple theory of a two-state system is 
%that I don’t want to have trap frequencies or atomic quantities which are far removed from reality.}

The main features are as follows.  At very low $\wtrap$ the value of $P$ tends to $6.15\times 10^5\,{\rm m}^{-1}$, the value given by the quasi-static theory (equation (\ref{T})).  As $\wtrap$ increases, there is an initial drop; the
population of the upper state has a time constant of $1/A_{21}$ and cannot remain in equilibrium with the changing radiation field.  In the frequency domain, the separation of the Fourier components of the radiation field experienced by the ion increases until only the carrier lies under the response curve.  There is a corresponding initial increase in the out-of-phase term $Q$, but this amplitude also then falls off as the damping effect of the slow spontaneous emission becomes more marked.  However, this behaviour is superposed on a resonant feature, reminiscent of a simple harmonic oscillator.  A two-state system can exhibit Rabi oscillations at the frequency
$\Omega_{12}' = (\Omega_{12}^2+\Delta_{12}^2)^{1/2}$,
i.e., $20.62\Mrads$ in this case.  The populations can then change rapidly on a time-scale much shorter than that of spontaneous decay because the effect is entirely due to stimulated processes.  Normally, in an equilibrium situation, any such oscillations have been damped out by the randomizing effect of spontaneous emission. However, in the present case the changes in the radiation field give rise to forced oscillations which grow in amplitude
as $\wtrap \rightarrow \Omega_{12}'$.  Of course, spontaneous emission---on which the cooling process
depends---still occurs; the analysis thus suggests that one can get enhanced cooling as one approaches the resonance, as shown in the figure.  It should nevertheless be pointed out that this phenomenon is not universally exploitable.  The parameters on which figure 1 is based were chosen to illustrate the two distinct physical effects which contribute to the curve, requiring $A_{21}$ to be significantly less than $\Omega_{12}'$, and hence also less than the trap frequency on resonance.  This condition would not be fulfilled for the strong transitions normally used for cooling.  If it is not satisfied, the initial drop and the broadened resonant feature overlap, and there is no appreciable improvement in cooling.

The phenomenon which we have discussed in this section is entirely unrelated to the dipole force and does not 
concern an ion near the ground state of motion. It is therefore not the subject of such works as \cite{0022-3700-17-16-019,Wineland:92,PhysRevA.46.2668,PhysRevA.49.421}. On the other hand, the
phenomenon described in this section can be related to the one exploited for
quantum logic gates in the proposal of Jonathan {\em et al.} \cite{PhysRevA.62.042307,PhysRevLett.87.127901}. This relies
on the resonance which occurs when the a.c. Stark shift matches the motional frequency.

\section{The three-state system}  \label{s.three}

\begin{figure}[t!]
\begin{center}
\hspace{1cm}\begin{tikzpicture}[scale=1] \begin{axis}[
xmin=-1500, xmax=1500,
ymin=0, ymax=0.3,
xlabel={$\Delta_{32} / 10^6$ rad s$^{-1}$},
ylabel={Population of $|2\rangle$},
width=0.8 \textwidth, height=0.6 \textwidth, yticklabels={0,0,0.05,0.1,0.15,0.2,0.25,0.3},
]
\addplot [ultra thin]
coordinates {
(0,0)
(0,0.3)
};
\addplot [ ]
coordinates {
(-1500,0.00387783)
(-1494,0.00390895)
(-1488,0.00394045)
(-1482,0.00397233)
(-1476,0.0040046)
(-1470,0.00403726)
(-1464,0.00407032)
(-1458,0.00410379)
(-1452,0.00413768)
(-1446,0.00417199)
(-1440,0.00420672)
(-1434,0.00424189)
(-1428,0.0042775)
(-1422,0.00431356)
(-1416,0.00435008)
(-1410,0.00438706)
(-1404,0.00442452)
(-1398,0.00446246)
(-1392,0.00450089)
(-1386,0.00453982)
(-1380,0.00457925)
(-1374,0.0046192)
(-1368,0.00465967)
(-1362,0.00470068)
(-1356,0.00474222)
(-1350,0.00478433)
(-1344,0.00482699)
(-1338,0.00487023)
(-1332,0.00491405)
(-1326,0.00495846)
(-1320,0.00500348)
(-1314,0.00504911)
(-1308,0.00509537)
(-1302,0.00514227)
(-1296,0.00518982)
(-1290,0.00523803)
(-1284,0.00528691)
(-1278,0.00533648)
(-1272,0.00538675)
(-1266,0.00543773)
(-1260,0.00548944)
(-1254,0.00554189)
(-1248,0.00559509)
(-1242,0.00564906)
(-1236,0.00570382)
(-1230,0.00575937)
(-1224,0.00581574)
(-1218,0.00587294)
(-1212,0.00593098)
(-1206,0.00598989)
(-1200,0.00604968)
(-1194,0.00611037)
(-1188,0.00617198)
(-1182,0.00623452)
(-1176,0.00629801)
(-1170,0.00636248)
(-1164,0.00642794)
(-1158,0.00649441)
(-1152,0.00656193)
(-1146,0.00663049)
(-1140,0.00670013)
(-1134,0.00677088)
(-1128,0.00684275)
(-1122,0.00691577)
(-1116,0.00698997)
(-1110,0.00706536)
(-1104,0.00714197)
(-1098,0.00721984)
(-1092,0.00729899)
(-1086,0.00737944)
(-1080,0.00746122)
(-1074,0.00754437)
(-1068,0.00762892)
(-1062,0.0077149)
(-1056,0.00780233)
(-1050,0.00789125)
(-1044,0.0079817)
(-1038,0.00807372)
(-1032,0.00816732)
(-1026,0.00826256)
(-1020,0.00835948)
(-1014,0.00845811)
(-1008,0.00855849)
(-1002,0.00866066)
(-996,0.00876467)
(-990,0.00887056)
(-984,0.00897838)
(-978,0.00908817)
(-972,0.00919998)
(-966,0.00931387)
(-960,0.00942987)
(-954,0.00954806)
(-948,0.00966847)
(-942,0.00979117)
(-936,0.00991623)
(-930,0.0100437)
(-924,0.0101736)
(-918,0.010306)
(-912,0.0104411)
(-906,0.0105788)
(-900,0.0107193)
(-894,0.0108625)
(-888,0.0110086)
(-882,0.0111577)
(-876,0.0113099)
(-870,0.0114652)
(-864,0.0116236)
(-858,0.0117854)
(-852,0.0119506)
(-846,0.0121192)
(-840,0.0122915)
(-834,0.0124674)
(-828,0.0126471)
(-822,0.0128307)
(-816,0.0130184)
(-810,0.0132101)
(-804,0.0134061)
(-798,0.0136065)
(-792,0.0138114)
(-786,0.014021)
(-780,0.0142354)
(-774,0.0144546)
(-768,0.014679)
(-762,0.0149086)
(-756,0.0151437)
(-750,0.0153843)
(-744,0.0156306)
(-738,0.0158829)
(-732,0.0161414)
(-726,0.0164061)
(-720,0.0166774)
(-714,0.0169555)
(-708,0.0172405)
(-702,0.0175328)
(-696,0.0178325)
(-690,0.01814)
(-684,0.0184554)
(-678,0.018779)
(-672,0.0191113)
(-666,0.0194523)
(-660,0.0198025)
(-654,0.0201623)
(-648,0.0205318)
(-642,0.0209115)
(-636,0.0213019)
(-630,0.0217031)
(-624,0.0221158)
(-618,0.0225402)
(-612,0.0229769)
(-606,0.0234264)
(-600,0.023889)
(-594,0.0243655)
(-588,0.0248561)
(-582,0.0253617)
(-576,0.0258827)
(-570,0.0264198)
(-564,0.0269736)
(-558,0.0275449)
(-552,0.0281344)
(-546,0.0287428)
(-540,0.029371)
(-534,0.0300199)
(-528,0.0306902)
(-522,0.031383)
(-516,0.0320992)
(-510,0.03284)
(-504,0.0336064)
(-498,0.0343996)
(-492,0.0352209)
(-486,0.0360716)
(-480,0.036953)
(-474,0.0378667)
(-468,0.0388141)
(-462,0.0397971)
(-456,0.0408172)
(-450,0.0418763)
(-444,0.0429765)
(-438,0.0441198)
(-432,0.0453085)
(-426,0.0465448)
(-420,0.0478313)
(-414,0.0491707)
(-408,0.0505658)
(-402,0.0520196)
(-396,0.0535353)
(-390,0.0551163)
(-384,0.0567662)
(-378,0.0584891)
(-372,0.0602888)
(-366,0.0621699)
(-360,0.0641371)
(-354,0.0661952)
(-348,0.0683496)
(-342,0.0706059)
(-336,0.0729701)
(-330,0.0754485)
(-324,0.0780479)
(-318,0.0807754)
(-312,0.0836385)
(-306,0.0866451)
(-300,0.0898033)
(-294,0.093122)
(-288,0.09661)
(-282,0.100277)
(-276,0.104131)
(-270,0.108183)
(-264,0.112441)
(-258,0.116914)
(-252,0.121611)
(-246,0.126538)
(-240,0.131701)
(-234,0.137103)
(-228,0.142742)
(-222,0.148614)
(-216,0.154707)
(-210,0.161001)
(-204,0.167466)
(-198,0.17406)
(-192,0.180721)
(-186,0.187367)
(-180,0.193892)
(-174,0.200159)
(-168,0.205994)
(-162,0.211187)
(-156,0.215488)
(-150,0.218609)
(-144,0.220231)
(-138,0.220025)
(-132,0.21767)
(-126,0.212895)
(-120,0.205515)
(-114,0.195479)
(-108,0.182905)
(-102,0.168102)
(-96,0.151561)
(-90,0.133921)
(-84,0.115915)
(-78,0.0982898)
(-72,0.0817471)
(-66,0.0668819)
(-60,0.054155)
(-54,0.0438835)
(-48,0.0362508)
(-42,0.0313279)
(-36,0.0290984)
(-30,0.0294832)
(-24,0.0323604)
(-18,0.0375796)
(-12,0.0449707)
(-6,0.0543481)
(0,0.0655102)
(6,0.078237)
(12,0.0922867)
(18,0.107392)
(24,0.12326)
(30,0.139568)
(36,0.155975)
(42,0.172124)
(48,0.187656)
(54,0.202227)
(60,0.215523)
(66,0.227274)
(72,0.237274)
(78,0.245383)
(84,0.251533)
(90,0.25573)
(96,0.258041)
(102,0.258588)
(108,0.257531)
(114,0.255056)
(120,0.251363)
(126,0.246654)
(132,0.241123)
(138,0.234951)
(144,0.228302)
(150,0.221321)
(156,0.214133)
(162,0.206842)
(168,0.199535)
(174,0.192282)
(180,0.18514)
(186,0.17815)
(192,0.171347)
(198,0.164753)
(204,0.158384)
(210,0.152251)
(216,0.146358)
(222,0.140707)
(228,0.135297)
(234,0.130123)
(240,0.125181)
(246,0.120463)
(252,0.115962)
(258,0.111669)
(264,0.107577)
(270,0.103675)
(276,0.0999565)
(282,0.0964115)
(288,0.0930317)
(294,0.089809)
(300,0.0867355)
(306,0.0838034)
(312,0.0810054)
(318,0.0783345)
(324,0.0757841)
(330,0.0733477)
(336,0.0710194)
(342,0.0687936)
(348,0.0666647)
(354,0.0646278)
(360,0.0626779)
(366,0.0608107)
(372,0.0590217)
(378,0.057307)
(384,0.0556628)
(390,0.0540855)
(396,0.0525717)
(402,0.0511181)
(408,0.049722)
(414,0.0483802)
(420,0.0470903)
(426,0.0458497)
(432,0.044656)
(438,0.043507)
(444,0.0424005)
(450,0.0413346)
(456,0.0403073)
(462,0.0393169)
(468,0.0383617)
(474,0.0374401)
(480,0.0365506)
(486,0.0356917)
(492,0.0348621)
(498,0.0340605)
(504,0.0332857)
(510,0.0325365)
(516,0.0318119)
(522,0.0311109)
(528,0.0304323)
(534,0.0297753)
(540,0.0291391)
(546,0.0285227)
(552,0.0279254)
(558,0.0273463)
(564,0.0267848)
(570,0.0262403)
(576,0.0257119)
(582,0.0251991)
(588,0.0247013)
(594,0.0242179)
(600,0.0237484)
(606,0.0232923)
(612,0.022849)
(618,0.0224182)
(624,0.0219992)
(630,0.0215918)
(636,0.0211955)
(642,0.0208099)
(648,0.0204346)
(654,0.0200692)
(660,0.0197135)
(666,0.0193671)
(672,0.0190296)
(678,0.0187009)
(684,0.0183805)
(690,0.0180681)
(696,0.0177636)
(702,0.0174667)
(708,0.0171771)
(714,0.0168946)
(720,0.016619)
(726,0.01635)
(732,0.0160874)
(738,0.0158311)
(744,0.0155808)
(750,0.0153363)
(756,0.0150976)
(762,0.0148643)
(768,0.0146363)
(774,0.0144135)
(780,0.0141958)
(786,0.0139829)
(792,0.0137747)
(798,0.0135711)
(804,0.013372)
(810,0.0131772)
(816,0.0129866)
(822,0.0128001)
(828,0.0126175)
(834,0.0124388)
(840,0.0122639)
(846,0.0120926)
(852,0.0119248)
(858,0.0117605)
(864,0.0115996)
(870,0.0114419)
(876,0.0112874)
(882,0.011136)
(888,0.0109876)
(894,0.0108422)
(900,0.0106996)
(906,0.0105598)
(912,0.0104227)
(918,0.0102882)
(924,0.0101563)
(930,0.0100269)
(936,0.0099)
(942,0.00977546)
(948,0.00965324)
(954,0.00953329)
(960,0.00941555)
(966,0.00929998)
(972,0.00918651)
(978,0.0090751)
(984,0.00896569)
(990,0.00885825)
(996,0.00875272)
(1002,0.00864906)
(1008,0.00854722)
(1014,0.00844717)
(1020,0.00834885)
(1026,0.00825224)
(1032,0.00815729)
(1038,0.00806396)
(1044,0.00797223)
(1050,0.00788204)
(1056,0.00779337)
(1062,0.00770618)
(1068,0.00762045)
(1074,0.00753614)
(1080,0.00745321)
(1086,0.00737164)
(1092,0.0072914)
(1098,0.00721245)
(1104,0.00713478)
(1110,0.00705836)
(1116,0.00698315)
(1122,0.00690914)
(1128,0.00683629)
(1134,0.00676458)
(1140,0.006694)
(1146,0.00662451)
(1152,0.00655609)
(1158,0.00648873)
(1164,0.0064224)
(1170,0.00635708)
(1176,0.00629275)
(1182,0.00622938)
(1188,0.00616697)
(1194,0.00610549)
(1200,0.00604492)
(1206,0.00598525)
(1212,0.00592645)
(1218,0.00586851)
(1224,0.00581142)
(1230,0.00575515)
(1236,0.0056997)
(1242,0.00564504)
(1248,0.00559117)
(1254,0.00553806)
(1260,0.0054857)
(1266,0.00543408)
(1272,0.00538318)
(1278,0.00533299)
(1284,0.0052835)
(1290,0.0052347)
(1296,0.00518656)
(1302,0.00513909)
(1308,0.00509227)
(1314,0.00504608)
(1320,0.00500051)
(1326,0.00495556)
(1332,0.00491121)
(1338,0.00486745)
(1344,0.00482427)
(1350,0.00478167)
(1356,0.00473962)
(1362,0.00469813)
(1368,0.00465718)
(1374,0.00461676)
(1380,0.00457687)
(1386,0.00453748)
(1392,0.00449861)
(1398,0.00446023)
(1404,0.00442233)
(1410,0.00438492)
(1416,0.00434798)
(1422,0.00431151)
(1428,0.00427549)
(1434,0.00423992)
(1440,0.00420479)
(1446,0.0041701)
(1452,0.00413583)
(1458,0.00410198)
(1464,0.00406855)
(1470,0.00403552)
(1476,0.00400289)
(1482,0.00397066)
(1488,0.00393881)
(1494,0.00390734)
};
\end{axis} 
%\myBoundingBox
\end{tikzpicture}\\
\rule{1cm}{0pt}\begin{tikzpicture}[scale=1] \input{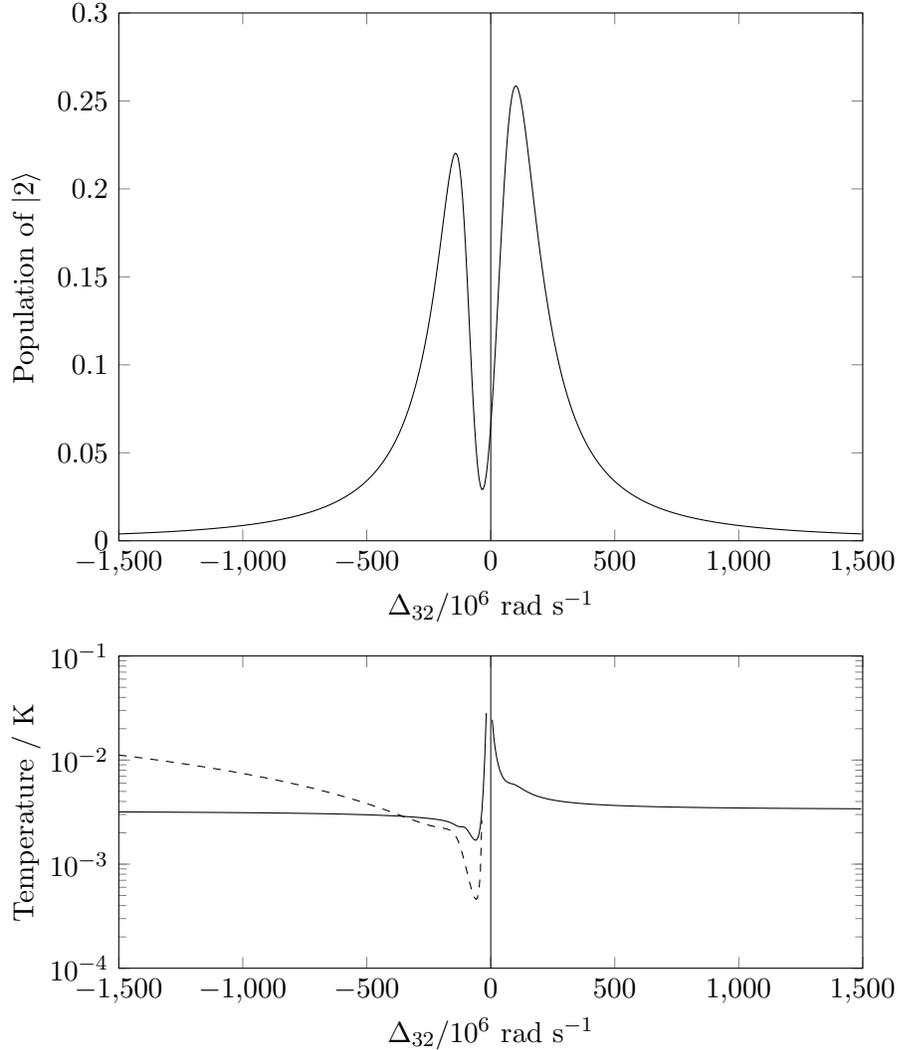} 
%\myBoundingBox
\end{tikzpicture}
\end{center}

\caption{The three state system.  The upper curve is the mean population of $\ket{2}$ as a function of the detuning $\Delta_{32}$ of the repumper. The lower plots give the corresponding equilibrium temperatures $T$ according to the quasi-static theory (dashed curve) and the dynamic theory (full curve).  The former is not plotted
for $\Delta_{32}$ higher than the minimum of the dark resonance because the theory predicts indefinite heating at these frequencies.  The dynamic theory predicts a high temperature on the high frequency side of the dark resonance but as the frequency increases cooling is quickly re-established.  See text for discussion and the values of the parameters.}
\end{figure}

We now turn to the three-state system, in which the cooling again takes place on the
transition $\ket{1}$--$\ket{2}$ at $\lambda_{12}$ but there is a weak decay 
from $\ket{2}$ to $\ket{3}$ at $\lambda_{32}$ \cite{Lindberg:86,PhysRevA.11.1641}. 
It is straightforward to apply the analysis above to any given set of parameters; for the purposes of illustration we take $\lambda_{12} = 400\,$nm, 
$\lambda_{32} = 800\,$nm, 
$A_{21} = 125\times 10^6\,{\rm s}^{-1}$,
$A_{23} = 10\times 10^6\,{\rm s}^{-1}$,
$\Omega_{12} = 200\Mrads$,
$\Omega_{32} = 50\Mrads$
and  $\Delta_{12} = -35\Mrads$.  With this value of $\Delta_{12}$ in a two-state system one would normally expect Doppler cooling to be effective.  For simplicity we work in the regime $\wtrap \ll A_{21}$ to exclude the two-state effects outlined above.  We are interested in how the behaviour of the system changes as $\wtrap$ increases; however, we first consider a particular trap frequency $\wtrap = 25\Mrads$, high enough for dynamic effects to be significant.  Figure 2 shows the equilibrium temperature as a function of $\Delta_{32}$ according to the quasi-static theory and the dynamic theory.  The upper state population for a stationary ion is also shown; the sharp minimum, a dark resonance, occurs at $\Delta_{12} = \Delta_{32}$.  We note that for the three-state system the laser linewidths play an important role around this minimum \cite{PhysRevA.70.053802}, and we set them both to the value of $5.0\Mrads$.  As the plot shows, the temperature according to the dynamic theory settles in both wings to a steady value, while the predictions of the quasi-static theory are quite different; in particular, in the blue wing the ion does not cool at all.

\begin{figure} [t!]
\begin{center}
\hspace{1cm}\begin{tikzpicture}[scale=1] \input{fig3} 
%\myBoundingBox
\end{tikzpicture}\\
\rule{1cm}{0pt}\begin{tikzpicture}[scale=1] \input{fig3b} 
%\myBoundingBox
\end{tikzpicture}
\end{center}
\caption{Amplitudes of the in-phase fluctuations $u_1,\,u_2,\,u_3$ for repumper detunings of (i) $-1500\Mrads$; (ii) $+1500\Mrads$.  At very low frequency, the population in both cases mainly swings backwards and forwards between the states $\ket{1}$ and $\ket{3}$, but the small amplitude $u_2$ is negative in (ii) leading to heating,
as predicted by the quasi-static theory. However, for $\wtrap \gtrsim A_{23}$ the population of $\ket{3}$ is unable to follow the rapidly changing radiation field and becomes constant, so $u_3 \rightarrow 0$.  The remaining population then oscillates between $\ket{1}$ and $\ket{2}$ in phase with the motion in the trap as it would for a two-state system (provided that dynamic effects in the $\ket{1}$--$\ket{2}$  system can be neglected).  Parameters as for figure~2.}
\end{figure}

We explain this different behaviour by considering how the in-phase amplitudes of the components of the state populations $u_1,\; u_2,\; u_3$ change with $\wtrap$.  We distinguish three regions in figure 2: (i) large red detuning (ii) large blue detuning and (iii) $\Delta_{32} \sim \Delta_{12}$.  Region (iii) is complicated by coherent effects, so we first consider the other cases.  According to the quasi-static theory, the cooling process is less effective in (ii) than in (i) because in (i) the repumping rate is higher when $V_x$ is negative than when it is positive, tending to increase $u_2$.   The converse is true in (ii), so that in some circumstances $u_2$ can be negative.
Negative $u_2$ is here owing to the fact that when the repumper is blue-detuned, the repumping process increases the scattering when the ion's velocity is positive, with the result that the ion is accelerated on average, even though the main cooling laser remains red-detuned.

\begin{figure} [t!]
\begin{center}
\hspace{1cm}\begin{tikzpicture}[scale=1] \begin{axis}[
xmin=-250, xmax=0,
ymin=0, ymax=0.3,
xlabel={$\Delta_{32} / 10^6$ rad s$^{-1}$},
ylabel={Population of $|2\rangle$},
width=0.7 \textwidth, height=0.5 \textwidth, yticklabels={0,0,0.05,0.1,0.15,0.2,0.25,0.3},
]
\addplot []
coordinates {
(-250,0.123228)
(-249.5,0.123636)
(-249,0.124046)
(-248.5,0.124457)
(-248,0.12487)
(-247.5,0.125285)
(-247,0.125701)
(-246.5,0.126119)
(-246,0.126538)
(-245.5,0.12696)
(-245,0.127382)
(-244.5,0.127807)
(-244,0.128233)
(-243.5,0.128661)
(-243,0.12909)
(-242.5,0.129521)
(-242,0.129954)
(-241.5,0.130388)
(-241,0.130824)
(-240.5,0.131262)
(-240,0.131701)
(-239.5,0.132142)
(-239,0.132585)
(-238.5,0.133029)
(-238,0.133475)
(-237.5,0.133923)
(-237,0.134372)
(-236.5,0.134823)
(-236,0.135276)
(-235.5,0.13573)
(-235,0.136186)
(-234.5,0.136644)
(-234,0.137103)
(-233.5,0.137564)
(-233,0.138026)
(-232.5,0.13849)
(-232,0.138956)
(-231.5,0.139424)
(-231,0.139893)
(-230.5,0.140364)
(-230,0.140836)
(-229.5,0.14131)
(-229,0.141786)
(-228.5,0.142263)
(-228,0.142742)
(-227.5,0.143223)
(-227,0.143705)
(-226.5,0.144188)
(-226,0.144674)
(-225.5,0.145161)
(-225,0.145649)
(-224.5,0.146139)
(-224,0.146631)
(-223.5,0.147124)
(-223,0.147619)
(-222.5,0.148116)
(-222,0.148614)
(-221.5,0.149113)
(-221,0.149614)
(-220.5,0.150117)
(-220,0.150621)
(-219.5,0.151126)
(-219,0.151633)
(-218.5,0.152142)
(-218,0.152652)
(-217.5,0.153164)
(-217,0.153676)
(-216.5,0.154191)
(-216,0.154707)
(-215.5,0.155224)
(-215,0.155742)
(-214.5,0.156262)
(-214,0.156783)
(-213.5,0.157306)
(-213,0.15783)
(-212.5,0.158355)
(-212,0.158882)
(-211.5,0.15941)
(-211,0.159939)
(-210.5,0.160469)
(-210,0.161001)
(-209.5,0.161534)
(-209,0.162068)
(-208.5,0.162603)
(-208,0.163139)
(-207.5,0.163676)
(-207,0.164215)
(-206.5,0.164754)
(-206,0.165295)
(-205.5,0.165836)
(-205,0.166379)
(-204.5,0.166922)
(-204,0.167466)
(-203.5,0.168012)
(-203,0.168558)
(-202.5,0.169105)
(-202,0.169653)
(-201.5,0.170201)
(-201,0.170751)
(-200.5,0.1713)
(-200,0.171851)
(-199.5,0.172403)
(-199,0.172955)
(-198.5,0.173507)
(-198,0.17406)
(-197.5,0.174613)
(-197,0.175167)
(-196.5,0.175722)
(-196,0.176276)
(-195.5,0.176831)
(-195,0.177387)
(-194.5,0.177942)
(-194,0.178498)
(-193.5,0.179053)
(-193,0.179609)
(-192.5,0.180165)
(-192,0.180721)
(-191.5,0.181277)
(-191,0.181832)
(-190.5,0.182387)
(-190,0.182943)
(-189.5,0.183497)
(-189,0.184052)
(-188.5,0.184606)
(-188,0.185159)
(-187.5,0.185712)
(-187,0.186265)
(-186.5,0.186816)
(-186,0.187367)
(-185.5,0.187917)
(-185,0.188467)
(-184.5,0.189015)
(-184,0.189562)
(-183.5,0.190108)
(-183,0.190653)
(-182.5,0.191196)
(-182,0.191739)
(-181.5,0.192279)
(-181,0.192819)
(-180.5,0.193356)
(-180,0.193892)
(-179.5,0.194427)
(-179,0.194959)
(-178.5,0.195489)
(-178,0.196018)
(-177.5,0.196544)
(-177,0.197068)
(-176.5,0.19759)
(-176,0.198109)
(-175.5,0.198625)
(-175,0.199139)
(-174.5,0.19965)
(-174,0.200159)
(-173.5,0.200664)
(-173,0.201166)
(-172.5,0.201666)
(-172,0.202161)
(-171.5,0.202654)
(-171,0.203143)
(-170.5,0.203628)
(-170,0.204109)
(-169.5,0.204586)
(-169,0.20506)
(-168.5,0.205529)
(-168,0.205994)
(-167.5,0.206455)
(-167,0.20691)
(-166.5,0.207362)
(-166,0.207808)
(-165.5,0.208249)
(-165,0.208686)
(-164.5,0.209117)
(-164,0.209542)
(-163.5,0.209962)
(-163,0.210376)
(-162.5,0.210785)
(-162,0.211187)
(-161.5,0.211584)
(-161,0.211974)
(-160.5,0.212357)
(-160,0.212734)
(-159.5,0.213104)
(-159,0.213467)
(-158.5,0.213822)
(-158,0.214171)
(-157.5,0.214512)
(-157,0.214845)
(-156.5,0.215171)
(-156,0.215488)
(-155.5,0.215798)
(-155,0.216099)
(-154.5,0.216391)
(-154,0.216675)
(-153.5,0.21695)
(-153,0.217216)
(-152.5,0.217472)
(-152,0.217719)
(-151.5,0.217956)
(-151,0.218184)
(-150.5,0.218402)
(-150,0.218609)
(-149.5,0.218806)
(-149,0.218992)
(-148.5,0.219168)
(-148,0.219332)
(-147.5,0.219486)
(-147,0.219628)
(-146.5,0.219759)
(-146,0.219877)
(-145.5,0.219984)
(-145,0.220079)
(-144.5,0.220161)
(-144,0.220231)
(-143.5,0.220289)
(-143,0.220333)
(-142.5,0.220364)
(-142,0.220382)
(-141.5,0.220387)
(-141,0.220377)
(-140.5,0.220354)
(-140,0.220317)
(-139.5,0.220266)
(-139,0.2202)
(-138.5,0.22012)
(-138,0.220025)
(-137.5,0.219915)
(-137,0.21979)
(-136.5,0.219649)
(-136,0.219494)
(-135.5,0.219322)
(-135,0.219135)
(-134.5,0.218932)
(-134,0.218712)
(-133.5,0.218476)
(-133,0.218224)
(-132.5,0.217956)
(-132,0.21767)
(-131.5,0.217368)
(-131,0.217049)
(-130.5,0.216713)
(-130,0.216359)
(-129.5,0.215988)
(-129,0.215599)
(-128.5,0.215193)
(-128,0.214769)
(-127.5,0.214328)
(-127,0.213868)
(-126.5,0.213391)
(-126,0.212895)
(-125.5,0.212381)
(-125,0.211849)
(-124.5,0.211299)
(-124,0.21073)
(-123.5,0.210143)
(-123,0.209537)
(-122.5,0.208913)
(-122,0.208271)
(-121.5,0.207609)
(-121,0.20693)
(-120.5,0.206232)
(-120,0.205515)
(-119.5,0.20478)
(-119,0.204026)
(-118.5,0.203253)
(-118,0.202462)
(-117.5,0.201653)
(-117,0.200826)
(-116.5,0.19998)
(-116,0.199116)
(-115.5,0.198233)
(-115,0.197333)
(-114.5,0.196415)
(-114,0.195479)
(-113.5,0.194525)
(-113,0.193553)
(-112.5,0.192565)
(-112,0.191558)
(-111.5,0.190535)
(-111,0.189494)
(-110.5,0.188437)
(-110,0.187363)
(-109.5,0.186273)
(-109,0.185166)
(-108.5,0.184044)
(-108,0.182905)
(-107.5,0.181751)
(-107,0.180582)
(-106.5,0.179397)
(-106,0.178198)
(-105.5,0.176984)
(-105,0.175756)
(-104.5,0.174513)
(-104,0.173257)
(-103.5,0.171988)
(-103,0.170705)
(-102.5,0.16941)
(-102,0.168102)
(-101.5,0.166782)
(-101,0.16545)
(-100.5,0.164107)
(-100,0.162752)
(-99.5,0.161387)
(-99,0.160012)
(-98.5,0.158626)
(-98,0.157231)
(-97.5,0.155826)
(-97,0.154413)
(-96.5,0.152991)
(-96,0.151561)
(-95.5,0.150123)
(-95,0.148678)
(-94.5,0.147226)
(-94,0.145768)
(-93.5,0.144304)
(-93,0.142834)
(-92.5,0.141359)
(-92,0.139879)
(-91.5,0.138395)
(-91,0.136907)
(-90.5,0.135416)
(-90,0.133921)
(-89.5,0.132424)
(-89,0.130925)
(-88.5,0.129425)
(-88,0.127922)
(-87.5,0.12642)
(-87,0.124917)
(-86.5,0.123413)
(-86,0.121911)
(-85.5,0.120409)
(-85,0.118909)
(-84.5,0.117411)
(-84,0.115915)
(-83.5,0.114421)
(-83,0.112931)
(-82.5,0.111444)
(-82,0.109961)
(-81.5,0.108483)
(-81,0.107009)
(-80.5,0.10554)
(-80,0.104077)
(-79.5,0.102621)
(-79,0.10117)
(-78.5,0.0997264)
(-78,0.0982898)
(-77.5,0.096861)
(-77,0.09544)
(-76.5,0.0940274)
(-76,0.0926235)
(-75.5,0.0912286)
(-75,0.0898431)
(-74.5,0.0884674)
(-74,0.0871018)
(-73.5,0.0857466)
(-73,0.0844023)
(-72.5,0.083069)
(-72,0.0817471)
(-71.5,0.080437)
(-71,0.0791389)
(-70.5,0.0778531)
(-70,0.0765799)
(-69.5,0.0753197)
(-69,0.0740727)
(-68.5,0.0728391)
(-68,0.0716192)
(-67.5,0.0704133)
(-67,0.0692216)
(-66.5,0.0680444)
(-66,0.0668819)
(-65.5,0.0657343)
(-65,0.0646019)
(-64.5,0.0634849)
(-64,0.0623834)
(-63.5,0.0612977)
(-63,0.060228)
(-62.5,0.0591745)
(-62,0.0581372)
(-61.5,0.0571165)
(-61,0.0561125)
(-60.5,0.0551253)
(-60,0.054155)
(-59.5,0.0532019)
(-59,0.0522661)
(-58.5,0.0513476)
(-58,0.0504467)
(-57.5,0.0495634)
(-57,0.0486979)
(-56.5,0.0478503)
(-56,0.0470206)
(-55.5,0.046209)
(-55,0.0454156)
(-54.5,0.0446404)
(-54,0.0438835)
(-53.5,0.043145)
(-53,0.042425)
(-52.5,0.0417234)
(-52,0.0410405)
(-51.5,0.0403762)
(-51,0.0397305)
(-50.5,0.0391036)
(-50,0.0384954)
(-49.5,0.037906)
(-49,0.0373354)
(-48.5,0.0367837)
(-48,0.0362508)
(-47.5,0.0357368)
(-47,0.0352416)
(-46.5,0.0347654)
(-46,0.034308)
(-45.5,0.0338695)
(-45,0.0334499)
(-44.5,0.0330492)
(-44,0.0326673)
(-43.5,0.0323042)
(-43,0.03196)
(-42.5,0.0316346)
(-42,0.0313279)
(-41.5,0.03104)
(-41,0.0307707)
(-40.5,0.0305202)
(-40,0.0302883)
(-39.5,0.030075)
(-39,0.0298802)
(-38.5,0.0297039)
(-38,0.0295461)
(-37.5,0.0294067)
(-37,0.0292857)
(-36.5,0.0291829)
(-36,0.0290984)
(-35.5,0.0290321)
(-35,0.0289839)
(-34.5,0.0289537)
(-34,0.0289415)
(-33.5,0.0289473)
(-33,0.0289708)
(-32.5,0.0290123)
(-32,0.0290713)
(-31.5,0.029148)
(-31,0.0292423)
(-30.5,0.0293541)
(-30,0.0294832)
(-29.5,0.0296297)
(-29,0.0297934)
(-28.5,0.0299744)
(-28,0.0301723)
(-27.5,0.0303873)
(-27,0.0306191)
(-26.5,0.0308678)
(-26,0.0311332)
(-25.5,0.0314153)
(-25,0.0317139)
(-24.5,0.032029)
(-24,0.0323604)
(-23.5,0.0327081)
(-23,0.033072)
(-22.5,0.033452)
(-22,0.0338479)
(-21.5,0.0342598)
(-21,0.0346875)
(-20.5,0.0351308)
(-20,0.0355898)
(-19.5,0.0360642)
(-19,0.0365541)
(-18.5,0.0370592)
(-18,0.0375796)
(-17.5,0.038115)
(-17,0.0386654)
(-16.5,0.0392307)
(-16,0.0398108)
(-15.5,0.0404056)
(-15,0.0410149)
(-14.5,0.0416387)
(-14,0.0422768)
(-13.5,0.0429292)
(-13,0.0435958)
(-12.5,0.0442763)
(-12,0.0449707)
(-11.5,0.045679)
(-11,0.046401)
(-10.5,0.0471365)
(-10,0.0478855)
(-9.5,0.0486478)
(-9,0.0494234)
(-8.5,0.0502121)
(-8,0.0510137)
(-7.5,0.0518283)
(-7,0.0526557)
(-6.5,0.0534956)
(-6,0.0543481)
(-5.5,0.0552131)
(-5,0.0560902)
(-4.5,0.0569796)
(-4,0.057881)
(-3.5,0.0587943)
(-3,0.0597193)
(-2.5,0.0606561)
(-2,0.0616044)
(-1.5,0.062564)
(-1,0.063535)
(-0.5,0.0645171)
};
\end{axis} 
%\myBoundingBox
\end{tikzpicture}\\
\rule{1cm}{0pt}\begin{tikzpicture}[scale=1] \input{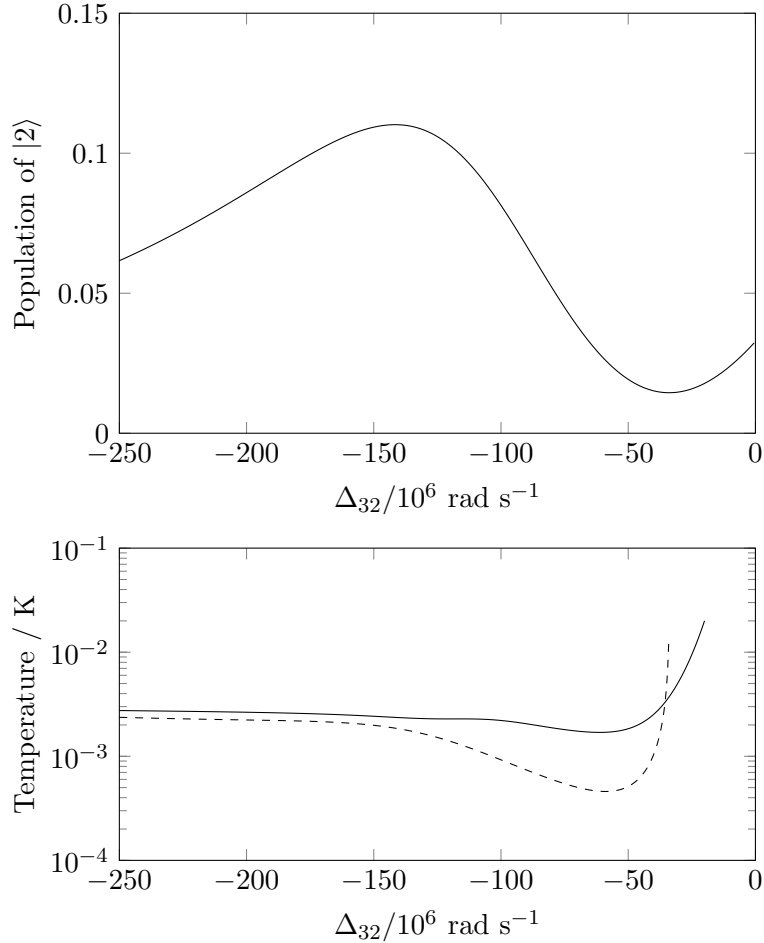} 
%\myBoundingBox
\end{tikzpicture}
\end{center}

\caption{Close-up of central region of figure 2, showing the predictions of the quasi-static theory and those of the dynamic theory with $\wtrap = 25\Mrads$ in region (iii) (see text).  The differences are less marked than in regions (i) and (ii) because $u_3$ does not tend to zero at high $\wtrap$ as coherent transfer does not involve spontaneous emission.}
\end{figure}

Indeed, for the conditions given, the quasi-static theory predicts that $u_2$ is always negative in (ii), so no cooling can occur.  The amplitudes $u_1$, $u_2$ and $u_3$ are shown as functions of $\wtrap$ in figure 3 
for $\Delta_{32} = \pm 1500\Mrads$.  In both cases the fact that the repumper is far from resonance causes there to be a significant fraction of the population in $\ket{3}$.  At very low $\wtrap$, this population oscillates (in antiphase with the population of $\ket{1}$) at a frequency $\wtrap$.  However, according to the dynamic theory, for trap frequencies too high for the spontaneous transition rate $\ket{2} \rightarrow \ket{3}$ to follow, $u_3 \rightarrow 0$, i.e., the population of $\ket{3}$ becomes constant and we have effectively a two-state system.  The equilibrium temperature becomes almost independent of $\wtrap$ in (i) and (ii) tending to the same value in both regions.

In region (iii), the differences in the predictions of the quasi-static and the dynamic theories are less marked; $u_3$ can be significant at any trap frequency because in the dark resonance region population can transfer coherently between $\ket{1}$ and $\ket{3}$ without spontaneous emission being involved.  At the minimum of the dark resonance, where $\Delta_{12} = \Delta_{32}$, the population of $\ket{2}$ for a stationary atom with lasers of zero linewidth would be zero and even for our conditions it is very small irrespective of $\wtrap$.  We show a close-up of the temperature plot for this region in figure 4.  The marked drop in temperature as the bottom of the dark resonance is approached from lower detuning can be understood from equation (\ref{TP}); a low value of $R_0$ occurs together with a large value of $P$.  This phenomenon has been exploited in cooling \Caft\ to below the Doppler limit \cite{1367-2630-18-2-023043}; see also \cite{PhysRevLett.85.4458,PhysRevLett.85.5547,PhysRevA.67.033402}.

\section{The ions {\boldmath \Caf} and {\boldmath \Caft}}  \label{s.ion}

\begin{figure} [t!]
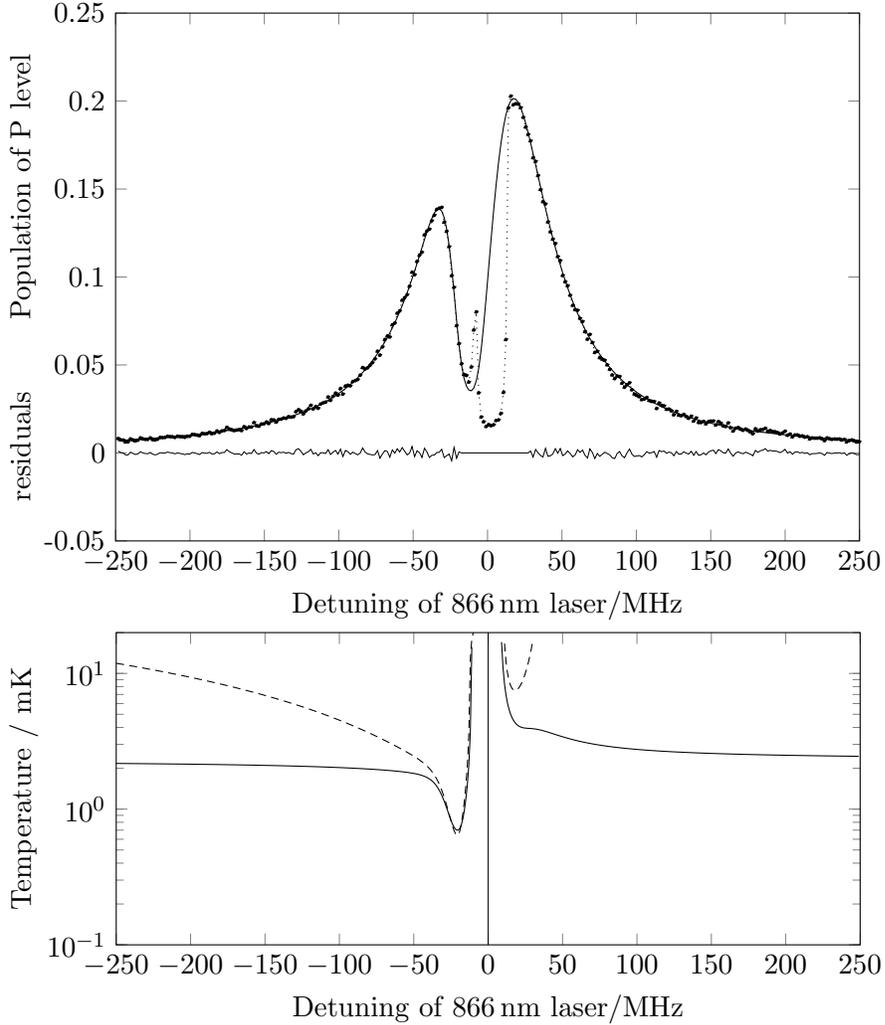

\begin{center}
\hspace{1cm}\begin{tikzpicture}[scale=1] \input{fig5} 
%\myBoundingBox
\end{tikzpicture}\\
\rule{1cm}{0pt}\begin{tikzpicture}[scale=1] \input{fig5b} 
%\myBoundingBox
\end{tikzpicture}
\end{center}

\caption{Upper plot: population of the 4\Phalf\ level of \Caf\ as a function of the frequency of the repumper laser at 866\,nm, as deduced from the observed fluorescence at 397nm.  Points and dotted line: experimental data.  The experimental parameters are given in the text.  Full curve: fit to the data, with a theory profile appropriate to a stationary ion.  Lower curve: residuals from best fit.  On the high frequency side of the dark resonance there is a region where the ion is hot, leading to a marked departure from the theoretical curve.  As the residual curve indicates, this region was omitted from the fit.  Lower plot: The equilibrium temperature as given by the quasi-static theory (dashed curve) and the dynamic theory (full curve).  For discussion see text.}
\end{figure}

We have described 2- and 3-state systems because of their simplicity, but one can use the same approach with multistate ions.  Indeed, it was the failure of the quasi-static theory to account for our observations on the alkali-like ions \Caf\ and \Caft\ that led to the development of the theory presented here.  As noted in 
the introduction, these ions have a cooling transition at $397\,$nm (4s\Shalf\ -- 4p\Phalf) with 
$A_{397} = 1.32\times 10^8\,{\rm s}^{-1}$, and there is a weak decay at $866\,$nm
(3d$^2{\rm D}_{3/2}$ -- 4p\Phalf) with
$A_{866} = 8.4\times 10^6\,{\rm s}^{-1}$.  We illustrate with a typical fluorescence spectrum observed from \Caf\ when the $866\,$nm laser is scanned (figure 5).  The theoretical profile used in the fit is that appropriate to a stationary ion; a region around the dark resonance has been omitted from the fit because of the evident distortion of the profile due to heating.  The trap frequency was 816 kHz, and the $B$-field was 1.5 gauss. 
The fitted parameters are: $397\,$nm detuning $-12.9\,$MHz, $397\,$nm intensity
$6.9 I_S$, $866\,$nm intensity $22.6 I_S$, where the saturation intensity
$I_S = (8\pi^3 \hbar\Gamma)/\lambda^3$ , with $\Gamma = A_{397} + A_{866}$, the FWHM of the 4\Phalf\ level. 
%The lasers are co-propagating; they are both linearly polarized perpendicular to the $B$-field 
%and have linewidths of order $1\,$MHz.  
The lasers are co-propagating and both are linearly polarized perpendicular to the $B$-field.
The model uses linewidths of order 1 MHz per beam.
The system behaves very similarly to the 3-state case; indeed, the differences between figures 2 and 5 are mainly due to the different experimental parameters.  On the low frequency side of the dark resonance, both theories predict cooling, though over most of this range the dynamic theory suggests a lower temperature.  On the steep rise out of the dark resonance on the high frequency side, both theories predict indefinite heating.  According to the dynamic theory, there is then a rapid recovery at higher frequencies.  The quasi-static theory predicts indefinite heating after a short region of moderate cooling, clearly at odds with the experimental observations.  In particular, if the quasi-static theory were a correct description, there would be virtually no fluorescence at frequencies above the dark resonance.  The fit to the experimental profile is much more consistent with the dynamic theory. 

We obtained an added confirmation of this interpretation by implementing a temperature diagnostic in the experiment. 
This consisted in observing the image of the trapped ion fluorescence on a CCD camera. The image became blurred in 
the region of the bottom of the dark resonance, indicating a high amplitude oscillation of the ion, and therefore a high temperature. The image became sharp again in both wings of the 866 nm laser scan, indicating a
low temperature in both cases.

A more quantitative experimental study using the \Caft\ ion has been reported \cite{1367-2630-18-2-023043}, in which the temperature of the ion was measured as a function of the various experimental parameters.  The aim was to take advantage of a dark resonance to Doppler-cool to a temperature low enough for further side-band cooling to take the ion to the ground state of the trap.  The results are discussed in \cite{HugoThesis}. We do not pursue this case further here, since there are complicating features which require detailed discussion: \Caft\ has a far more complex system of internal states than \Caf, the ion was in a $B$-field of 146 gauss (required to realise a clock transition for quantum information processing) and three laser beams were needed to prevent optical pumping to inaccessible states.  However, the dynamic theory is an essential feature of the theoretical treatment and the study gives further evidence of its validity.  

\section{Conclusion}

In this paper we have presented a simple technique which allows the response time of an atom or ion being Doppler-cooled to be incorporated into the theory of the cooling process.  We have discussed the results for two- and three-state systems, and showed that experimental data for \Caf\ which is not explicable on the basis of quasi-static theory is readily understood using the new approach. 

%\section*{Acknowledgements}

This work was funded by EPSRC.
We are very glad that this contribution can be included in a special issue for Prof Danny Segal. 
His good nature, irrepressible cheerfulness and ``can do" attitude were an inspiration to students and colleagues alike.

%\section*{Funding}
\vspace{12pt}

\bibliographystyle{tfp}
\bibliography{myrefs}

\end{document}